\documentclass[aps,floats,amsmath,amssymb,twocolumn,superscriptaddress,preprintnumbers,nofootinbib]{revtex4-1}
\pdfoutput=1
\usepackage{graphicx}
\usepackage{dcolumn}
\usepackage{bm}

\newcommand{\be}{\begin{eqnarray}}
\newcommand{\ee}{\end{eqnarray}}
\newcommand{\bea}{\begin{eqnarray}}
\newcommand{\eea}{\end{eqnarray}}

\newcommand{\GeV}{{~\rm GeV}}
\newcommand{\TeV}{{~\rm TeV}}

\newcommand{\MZZ}{M_Z}
\newcommand{\MW}{M_{W}}
\newcommand{\MHpm}{M_{{H^\pm}}}
\newcommand{\MH}{M_{H}}
\newcommand{\Mh}{M_{h}}

\newcommand{\MZp}{M_{Z'}}
\newcommand{\ZZ}{Z}
\newcommand{\Zp}{Z'}

\newcommand{\QQi}{Q_{Q_i}}
\newcommand{\Qui}{Q_{u_i}}
\newcommand{\Qdi}{Q_{d_i}}
\newcommand{\QHu}{Q_{H_u}}
\newcommand{\QHd}{Q_{H_d}}
\newcommand{\Qs}{Q_s}

\newcommand{\shad}{\sigma_{\rm had}}
\newcommand{\Uonep}{U(1)'}

\begin{document}

\title{A Higgsophilic $s$-channel $Z'$ and the CDF $W+2J$ Anomaly}
\author{JiJi Fan}
\email{jijifan@princeton.edu}
\affiliation{Department of Physics, Princeton University, Princeton, NJ 08544}
\author{David Krohn}
\email{dkrohn@physics.harvard.edu}
\affiliation{Department of Physics, Harvard University, Cambridge MA, 02138}
\author{Paul Langacker}
\email{pgl@ias.edu}
\affiliation{Department of Physics, Princeton University, Princeton, NJ 08544}
\affiliation{School of Natural Sciences, Institute for Advanced Study, Princeton, NJ 08540}
\author{Itay Yavin}
\email{iy5@nyu.edu}
\affiliation{Center for Cosmology and Particle Physics, Department of Physics, New York University, New York, NY 10003}

\date{\today}

\begin{abstract}
The CDF collaboration recently presented evidence for an excess in the dijet invariant mass distribution coming from events in a $W+2j$ exclusive sample.
Here we show that this excess can be explained by the $s$-channel production of a  weakly coupled Higgsophilic 
$\Zp$ near $\MZp \sim 270~{\rm GeV}$ which decays into a $W^\pm$ and a charged Higgs at $\MHpm \sim 150\GeV$. 
While the simplest  implementations of a general leptophobic $\Zp$ model quickly run into tensions with electroweak observables, a more specific Higgsophilic model evades these constraints without resorting to any fine-tuning.
We discuss the distinctive features of this model, focusing on its particular signatures at the Tevatron.
\end{abstract}

\maketitle

\section{Introduction}

The CDF collaboration recently presented an anomalous $3.2\sigma$ excess near $M_{jj}\sim150~{\rm GeV}$ in the dijet invariant mass distribution for events in a $W+2j$ exclusivesample~\cite{Aaltonen:2011mk}.  The source of this excess is as of yet
unclear, but given the difficult systematics involved and the fact that D0 did not observe a similar effect~\cite{Abazov:2011af}, the most plausible explanation seems to be a mismodeling of the  relevant backgrounds. However, a definitive Standard Model (SM) resolution of this question does not seem to be immediately forthcoming.  Indeed, CDF recently released preliminary results making use of a larger $7.3~{\rm fb}^{-1}$ sample and found that the statistical significance of the excess increases to nearly $5\sigma$.  Moreover, the collaboration checked several aspects of their background models and found that none was sufficient to account  for the excess.

Concurrently with the experimentalists' efforts to check all the systematics involved, it can prove useful to construct viable and predictive models of new physics that could account for the anomaly. These can serve as hypotheses to test and rule-out, and form an important part of the slow process towards a clearer picture. To this end, the theory community has been busy constructing explanatory models.  The field at this point is already crowded, with a series of models proposing that the excess can be accounted for by 
the associated production of a new resonance~\cite{Liu:2011di,Chen:2011wp,Enkhbat:2011qz,Kim:2011xv,Chang:2011wj,Babu:2011yw,Jung:2011ue,Fox:2011qd,Ko:2011ns,Buckley:2011vs,Jung:2011ua,Anchordoqui:2011ag,He:2011ss,Wang:2011ta,Nelson:2011us,Cheung:2011zt,Wang:2011uq,Yu:2011cw,Buckley:2011vc}, some means of $s$-channel production through a new field~\cite{Segre:2011tn,Carpenter:2011yj,Cao:2011yt,Eichten:2011sh,Kilic:2011sr,Wang:2011uq,nima}, subtleties in the treatment of SM physics~\cite{Campbell:2011gp,Plehn:2011nx,Sullivan:2011hu,He:2011ss}, or some other, more exotic explanations~\cite{Dobrescu:2011px,Sato:2011ui}. 

Most, but not all, of these models were constructed under the assumption that the CDF data disfavors the $s$-channel production mode. However, the more precise statement is that the data neither favors nor disfavors the existence of such a resonance\footnote{The preliminary analysis presented in~\cite{cdf73} using the full 7.3$fb^{-1}$ of data together with a more restricted set of cuts in fact seems to give some indication for the existence of such a resonance.}. In addition, it was usually assumed that the excess seen as the dijet resonance did not contain a significant concentration of heavy flavor. However, this inference was based on low statistics and depends upon whether one is looking at one $b$-tag or two $b$-tagged samples, so it is important to realize that these assumptions are not very strongly supported by data\footnote{Indeed, as we shall see below, the new data even seem to confirm departures from these assumptions.}. In the spirit of constructing testable hypotheses it is important that we do not bias ourselves too early and keep an open mind about the possible interpretation of the excess.

\begin{figure}[h]
\begin{center}
\includegraphics[width=0.4 \textwidth,height=0.2\textheight]{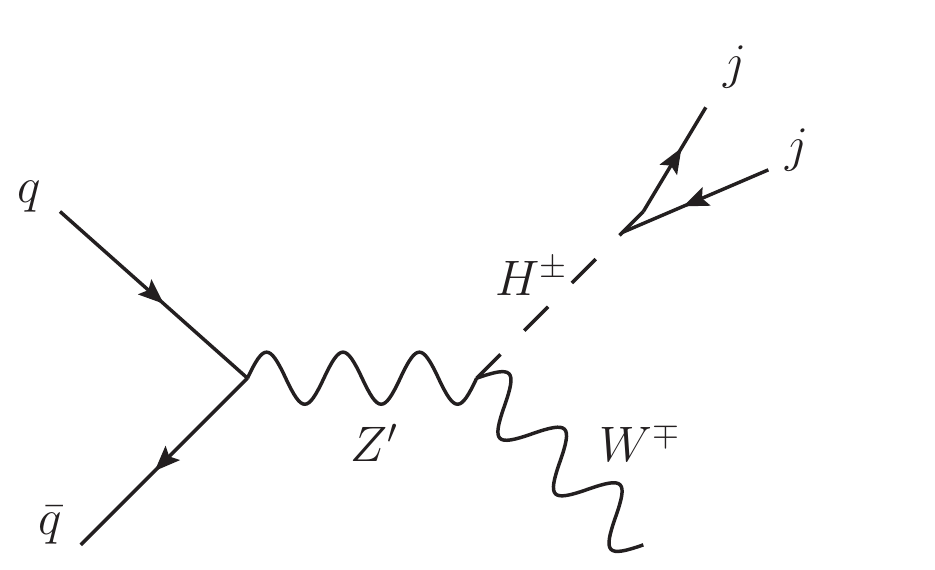}
\end{center}
\caption{The production and decay topology considered in this paper to explain the excess observed by the CDF collaboration.}
\label{fig:W+jj}
\end{figure}

Indeed, much can be had by allowing an $s$-channel resonance\footnote{We note that the authors of ref.~\cite{Eichten:2011sh} were the first to propose this topology as 
an explanation for the anomaly, and we thank Adam Martin for many  helpful discussions regarding the model.}. This possibility becomes particularly appealing when we realize that generically a $\Zp$ vector-boson couples to the $W^\pm H^\mp$ vertex, thereby manifesting the required signature shown in Fig.~\ref{fig:W+jj}, namely $p\bar{p}\rightarrow \Zp\rightarrow W^\pm(H^\mp\rightarrow jj)$. However, for most $\Zp$ models constructed outside the context of the CDF anomaly~\cite{Langacker:2008yv}, this decay mode is subdominant and one would expect  the $\Zp$ resonance to appear first in other channels\footnote{A similar mechanism for a leptophobic TeV-scale $ \Zp$ decay into Higgs and vector bosons was considered long ago in~\cite{Georgi:1996ei} and more recently in~\cite{Barger:2009xg}.}. In particular, $\Zp$ leptonic decays generally lead one to expect  $\MZp \gtrsim 800\GeV$. Fortunately, data is a strong antidote to prejudice and the CDF anomaly forces us to consider leptophobic $\Uonep$ models with a light $\Zp$ as a possibility. In contradistinction to previous $\Zp$ models employing a lighter $\MZp \sim 150\GeV$ boson produced in association with a $W$, here we focus on a heavier $\Zp$ produced in the $s$-channel. 
As we shall discuss in this paper, generic leptophobic models with $\MZp \sim 270\GeV$ manifesting the topology of Fig.~\ref{fig:W+jj} still exhibit some \textit{parametric} tensions with a variety of constraints, from Electroweak Precision Tests (EWPT) to dijet resonance searches. However, once recognized in their parametric form, these tensions point to a Higgsophilic $\Zp$ model as a viable explanation for the CDF excess. Moreover, through its contribution to the $\rho_0$ parameter, such a model also holds the hope of resolving the long-standing tension between the direct and indirect searches for the Higgs boson~\cite{Chanowitz:2002cd}.

The paper is organized as follows. In section \ref{sec:effectivemodels} we discuss leptophobic $\Zp$ models in general and present the associated formulae relevant for the phenomenology studied in this paper. In section \ref{sec:constraints} we review the constraints that place the strongest limits on such models. Section \ref{sec:wjjcdf} is devoted to the CDF anomaly and its possible interpretation in the context of leptophobic $s$-channel $\Zp$ models. In section \ref{sec:hemodels} we consider specific realizations of the general formalism, exhibit the tensions involved, and arrive at a viable Higgsophilic model as a possible explanation of the CDF anomaly. Finally, in section \ref{sec:discussion} we discuss possible tests of this hypothesis, suggest some future directions, and conclude.


\section{Effective Leptophobic Models}
\label{sec:effectivemodels}


In this section we present the relevant formulae and relations for a generic leptophobic model with general charge assignments. In the sections that follow we concentrate on several specific realizations of this general scheme. In particular, later we will show that Higgsophilic models constitute a viable explanation for the excess seen at CDF, consistent with all other constraints. 

\subsection{Charges and Mixing}

We denote the charge of the left handed quarks by $\QQi$, where $i=1,2,3$ denotes the generation. Similarly, $Q_{u_i^c}$, and $Q_{d_i^c}$ represent the charges of the conjugate fields, $u^c_i$ and $d^c_i$. In what follows we will concentrate on models with two Higgs doublet fields, $H_u$ and $H_d$, with charges $\QHu$, and $\QHd$, respectively. Since the $\Uonep$ is broken, we need not specify any particular relation among these charges even in the presence of Yukawa couplings between the Higgs fields and the fundamental matter fields. The completion to a full theory that exhibits the $\Uonep$ is left for the following sections. A coupling to the leptons is ultimately generated through mixing between the $\ZZ$ and $\Zp$ vector bosons; however, it is strongly suppressed and we leave the discussion of its effects to a separate section below. 

The appearance of $\Zp$ in the Higgs kinetic terms generate mixing with the SM bosons\footnote{For a review, see, e.g.,~\cite{Langacker:2008yv}.}. In particular, once the Higgs fields obtain a VEV, a mixing with the $\ZZ$ is present . The mass matrix is given by
\be
\nonumber
\mathcal{M}^2_{Z-Z'} =
\left(
\begin{array}{cc}
M_{Z^0}^2  & \Delta^2 \\
 &\\
 \Delta^2 & \MZp^2 
\end{array}
\right),
\ee
where
\be
M_{Z^0}^2 &=& \frac{1}{4}g_1^2\left(v_u^2 + v_d^2 \right) \\
\MZp^2 &=& g_2^2 \left( \QHu v_u^2 + \QHd v_d^2 + \Qs v_s^2 \right) \\
\Delta^2&=&\frac{1}{2}g_1g_2\left( \QHu v_u^2 - \QHd v_d^2 \right).
\ee
 $g_1^2 \equiv g^2 + g^{'2} = g^2/\cos^2\theta_{_W}$ is the $\ZZ$ coupling and $g_2$ is the $\Uonep$ coupling. $v_u = v \sin\beta$, $v_d = v\cos\beta$, and $v = 246\GeV$ are the VEV's of $\sqrt{2} H_u$ and $\sqrt{2} H_d$, respectively, and $v_s$ is the VEV of $\sqrt{2}S$, where $S$ is a SM-singlet field charged under 
 $\Uonep$. Since the mixing we will consider is much smaller than the mass difference, the mass eigenvalues are to a very good approximation given by
\be
M_1^2 &=& \MZZ^2 - \theta \Delta^2 \equiv M_Z^2 \\
M_2^2 &=& \MZp^2  + \theta \Delta^2 \sim  \MZp^2,
\ee
where the mixing angle is given exactly by 
\be
\theta = \frac{1}{2}\arctan\left(\frac{2\Delta^2}{\MZZ^2 - \MZp^2} \right) \approx \frac{\Delta^2}{\MZZ^2 - \MZp^2}.
\ee
This mixing has two important effects. First, the mass shift of the $\ZZ$ boson contributes to the $\rho_0$ parameter. Second, the mixing generates a coupling of $\Zp$ to the $\ZZ$ boson current at order $\theta$. We return to both effects below in section \ref{sec:EWPT}. 

\subsection{$\Zp$ Production}

The production of a $\Zp$ is dominated by first generation quark - anti-quark annihilation into the $\Zp$. At the Tevatron this proceeds mostly 
through the scattering of valence quarks, and
under this assumption we find the cross section is well described near $\MZp \sim 300\GeV$ by:
\be
\label{eqn:prodxsec}
\sigma_{\Zp} \approx g_2^2\left(\frac{300\GeV}{\MZp} \right)^{3.3} \left(\frac{5}{4}Q_{Q_1}^2 + Q_{u^c_1}^2 + \tfrac{1}{4}Q_{d^c_1}^2\right)
\ee
where $\sigma_{\Zp}$ is measured in nano-barns.

\subsection{$\Zp$ Decay}

For $m_{\Zp} < 2m_t$ the decay of the $Z'$ into jets is given by
\be
\label{eqn:Gammajj}
\frac{\Gamma_{jj}}{\MZp} = \frac{\alpha_2}{2}\left(\sum_{i=1,2,3}\left(\QQi^2+ Q_{d_i^c}^2\right) +\sum_{i=1,2}\left(\QQi^2+ Q_{u_i^c}^2\right)  \right),
\ee
where $\alpha_2\equiv g_2^2/4\pi$.
The decay into vector-bosons and Higgs bosons is given by~\cite{Georgi:1996ei}
\be
\label{eqn:GammaWH}
\frac{\Gamma_{W^\pm H^\mp}}{\MZp} &=& \frac{\alpha_2 }{6}\left( \QHu + \QHd \right)^2 \sin^2\beta\cos^2\beta
 \nonumber  \\
&\times& \xi\left((\MW/\MZp)^2, (\MHpm/\MZp)^2 \right);
\ee
\be
\label{eqn:GammaZH}
\frac{\Gamma_{Z H}}{\MZp} &=& \frac{\alpha_2}{12} \left( \QHu\sin\beta\sin\alpha - \QHd \cos\beta \cos\alpha  \right)^2  \nonumber \\
&\times& \xi\left((M_Z/\MZp)^2, (\MH/\MZp)^2 \right) \\
\frac{\Gamma_{Z h}}{\MZp} &=& \frac{\alpha_2}{12} \left(\QHu \sin\beta\cos\alpha + \QHd\cos\beta\sin\alpha \right)^2   \nonumber \\
&\times& \xi\left((M_Z/\MZp)^2,( \Mh/\MZp)^2 \right),
\ee
where $\Gamma_{W^\pm H^\mp}=\Gamma_{W^+ H^-}+\Gamma_{W^- H^+}$, $h$ and $H$ are the two CP-even Higgs scalars with mixing angle $\alpha$, and
\be
\nonumber
\xi(x,y) &=&  \left(1+ 2\left(5 x - y\right) + \left( x-y\right)^2\right)\lambda^{1/2}\left(1,x,y\right) \\ \nonumber
\lambda\left(x,y,z\right) &=&  x^2 + y^2 + z^2 - 2 x y - 2 y z - 2 z x.
\ee 
We have ignored $CP$ violation in the Higgs sector so there is no mixing with the pseudoscalar $A$.

We wrote Eqs.~(\ref{eqn:Gammajj}-\ref{eqn:GammaZH}) in an attempt to clearly exhibit the parametric tension between the decay into jets as compared with the decays into Higgs and vector bosons. Assuming no strong suppression from  phase-space factors, the ratio is given by
\be
\label{eqn:WHjjRatio}
\frac{\Gamma_{W^\pm H^\mp}}{\Gamma_{jj}} \approx \frac{1}{12} \frac{\left(\QHu+\QHd\right)^2}{\sum_q Q_q^2} \sin^22\beta.
\ee
In scenarios where the charges are all of order unity, this ratio is typically not more than $\sim 1/10$. As we will see momentarily, the decay of the charged Higgs requires one to be in the large $\tan\beta$ regime, where this ratio is further suppressed by approximately $\sin^22\beta$. Therefore, it is difficult to have a large enough cross-section to vector bosons without being in conflict with searches for resonances in the dijet sample. As we discuss in some more detail in section~\ref{sec:wjjcdf}, in order to explain the CDF anomaly one would require
\be
\sigma_{\Zp} \times {\rm BR}\left(\Zp \rightarrow W^\pm H^\mp \right) \approx 4{\rm~pb},
\ee
which, using Eq.~(\ref{eqn:WHjjRatio}) in the large $\tan\beta$ limit, implies a sizable contribution to the dijet sample of order
\be
\sigma_{\Zp} \times {\rm BR}\left(\Zp \rightarrow jj \right) \approx \frac{40{\rm~pb}  }{\sin^22\beta}.
\ee
This difficulty can be avoided if the relevant quark charges are small compared to those of the Higgs doublets.

Eqs.~(\ref{eqn:GammaWH}-\ref{eqn:GammaZH}) are valid in an arbitrary two Higgs doublet model. However,  we will mainly be interested in the case of moderate to large $\tan \beta$. We do not restrict ourselves to MSSM-type couplings, but comment here that in the MSSM\footnote{The $U(1)'$-extended MSSM does not allow elementary $\mu$ or $B \mu$ terms unless
$Q_{H_u}+Q_{H_d}=0$. An electroweak scale $\mu$ can be generated by an NMSSM~\cite{Ellwanger:2009dp}-like coupling
$\lambda_s S Q_{H_u} Q_{H_d}$. In general, the scalar component of $S$ can mix with the $h$ and $H$~\cite{Barger:2006dh}, but we ignore such mixing here. (The pseudoscalar component is eaten by the $Z'$.) } with large $\tan\beta$  and $M_{H^\pm}\sim 150$ GeV  one expects
$M_A$ and $M_H$  around 130 GeV and small $\alpha$. Small $\alpha$ and large $\tan \beta$ implies that
most of the symmetry breaking and the scalar $h$ are associated with $H_u$, while the fields in $H_d$ are close
in mass, even though we are not really in the decoupling limit. This region of parameters has the features that
the second Higgs doublet contributes very little to the electroweak oblique parameters; large supersymmetric loop contributions from the $\tilde t$ and $  t$ mainly affect $M_h$; and the $H\rightarrow W^+ W^-$ vertex is suppressed, 
so that the Tevatron searches are not sensitive to $H$. Furthermore, with the additional assumption $Q_{H_u}\ll Q_{H_d}$
(see below) one has that $\Gamma_{W^\pm H^\mp} \sim 2 \Gamma_{Z H} \gg \Gamma_{Z h}$.

Finally, we discuss the decay of $\Zp$ into leptons. Throughout we will concentrate on models where the leptons are uncharged under the $\Uonep$. Therefore, the only source of coupling between the mass eigenstate $\Zp$ and the leptons is through the mixing with the $\ZZ$ boson. Then the partial width into leptons is similar to that of the $\ZZ$ boson multiplied by the mixing angle,
\be
\label{eqn:PWlep}
\frac{\Gamma\left(\Zp\rightarrow l^+l^-\right)}{\MZp} = \frac{g^2\theta^2}{48\pi\cos^2\theta_{_W}}\left(\left| g_V\right|^2 + \left| g_A\right|^2 \right),
\ee
where $g_V = -\tfrac{1}{2}+2\sin^2\theta_{_W}$ and $g_A = -\tfrac{1}{2}$. The branching ratio is therefore approximately given by
\be
{\rm BR}\left(\Zp\rightarrow l^+l^-\right) \approx 0.28~ \theta^2 \left(\frac{\MZp}{300\GeV}\right)\left(\frac{1\GeV}{\Gamma_{\Zp}} \right).
\ee

\subsection{Higgs Decays}

The next phenomenological aspect we would like to discuss is the decay of the charged Higgs. It is rather independent of the details of the $\Uonep$ part of the model and has been worked out long ago (see e.g., ref.~\cite{Heinemeyer:2004gx,Djouadi:2005gj}). The two body decay into quarks is given by
\be\label{eqn:HpmTOcb1}
\frac{\Gamma(H^+ \rightarrow c \bar b)}{\MHpm}=\frac{3 |V_{cb}|^2}{8\pi \tan^2\beta} \left( |\kappa^{cb}_L|^2 +  |\kappa^{cb}_R|^2 \right), 
\ee
where
\be
\label{eqn:HpmTOcb}
\kappa^{cb}_L = \frac{ m_c}{v }, \qquad \kappa^{cb}_R = \frac{ m_b \tan^2\beta}{v},
\ee
and $v = 2 \MW/g = 246\GeV$. The same formula holds for the decay into a strange-charm pair, $\Gamma(H^+ \rightarrow c \bar s)$ except that $m_b\rightarrow m_s$ and  $V_{cb}\rightarrow V_{cs}\sim 1$. We note that in the large $\tan\beta$ limit, the decay into a bottom-charm pair increases with $\tan^2\beta$.

\begin{figure}[h]
\begin{center}
\includegraphics[width=0.4 \textwidth,height=0.2\textheight]{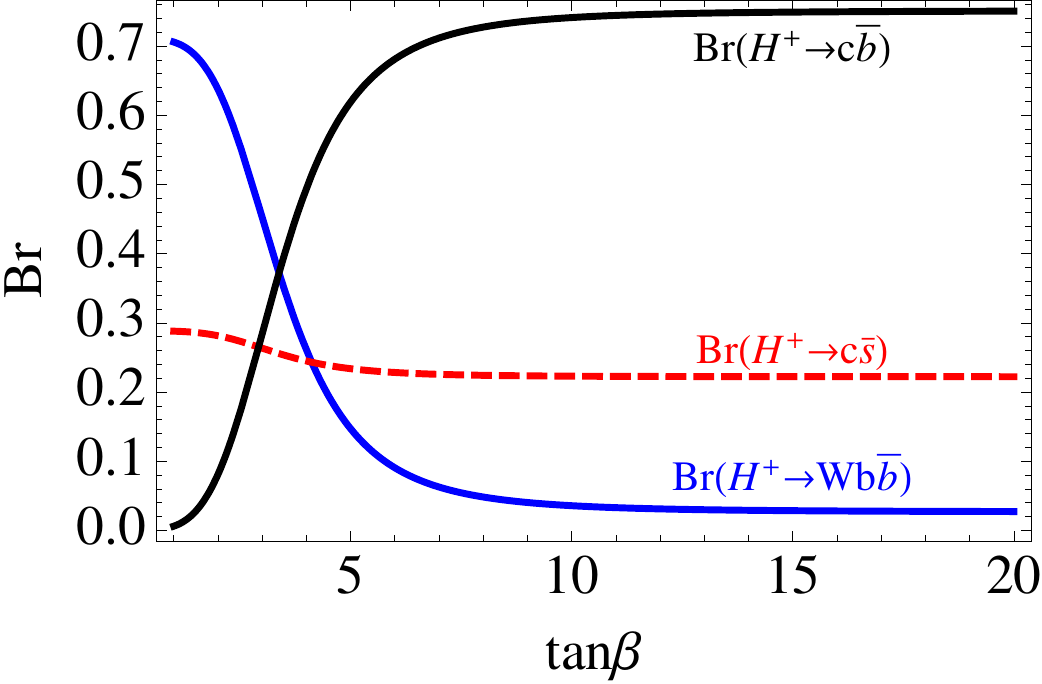}
\end{center}
\caption{The different branching ratios of the charged Higgs with $M_{H^+}=150$ GeV as a function of $\tan\beta$. The general behavior of these curves is not greatly affected by the mass of the charged Higgs in the range relevant for the CDF anomalous excess.}
\label{fig:HpmPW}
\end{figure}

Owing to the large top mass, the 3-body decay into two bottom quarks and a charged vector boson is in fact not negligible and is given by
\be
\label{eqn:HpmTObbW}
&&\frac{\Gamma\left(H^{\pm} \rightarrow W^\pm b\bar{b} \right)}{\MHpm} = \frac{3 }{128\pi^3 v^4 \tan^2\beta} \nonumber \\
&& \left(m_t^4 f_L\left(\kappa_t,\kappa_W\right)+\MHpm^2m_b^2\tan^4\beta f_R\left(\kappa_t,\kappa_W\right) \right). 
\ee
where the functions $f_L\left(\kappa_t,\kappa_W\right), f_R\left(\kappa_t,\kappa_W\right)$ are given in the appendix. If this type of model is to explain the CDF anomaly the charged Higgs boson should decay dominantly into two jets. Using the partial widths, Eqs.~(\ref{eqn:HpmTOcb1}) and~(\ref{eqn:HpmTObbW}), we find that the ratio between the 2-jet decay to the 3-body decay approximately scales as 
\be
\label{eqn:HpmPWRatio}
\frac{ \Gamma\left(H^{\pm} \rightarrow W^\pm b\bar{b} \right) }{ \Gamma\left(H^+ \rightarrow c \bar b\right) } \propto \tan^{-4}\beta.
\ee 
We therefore concentrate on the large $\tan\beta$ region throughout this paper. In Fig.~\ref{fig:HpmPW} we plot the exact branching ratios of the charged Higgs as a function of $\tan\beta$. 

\section{Constraints} 
\label{sec:constraints}


n this section we collect the most relevant constraints on leptophobic $\Zp$ models.

\subsection{Electroweak Precision Constraints}
\label{sec:EWPT}

The shift in the $\ZZ$ mass from $Z/\Zp$ mixing results in a shift of the $\rho_0$ parameter, where we follow the Particle Data Group convention~\cite{Nakamura:2010zzi} that $\delta \rho\equiv \rho_0-1$ represents the effects of new physics only, i.e., the effects of $m_t$ and $M_h$ are treated separately. The contribution to $\delta\rho$ is then
\be
\delta \rho = \frac{\Delta^4}{\MZZ^4}\frac{\MZZ^2}{\MZp^2 - \MZZ^2}.
\ee
The result of the global electroweak fit~\cite{Nakamura:2010zzi} is $\rho_0=1.0008^{+0.0017}_{-0.0007}$, while the corresponding Higgs mass is $M_h=162^{+265}_{-93}$ GeV
(the direct limits on $M_h$ from LEP~2 and the Tevatron are not included in this fit), with essentially no change in the other SM parameters with respect to the SM fit. What is happening is that the decrease in the predicted value of $M_Z$ due to 
$\rho_0 > 1$ is almost exactly compensated by the larger Higgs mass compared to the SM prediction of $90^{+27}_{-22}$ GeV. This contribution therefore eases the well-known tension\footnote{The implications of $\ZZ/\Zp$ mixing for the $M_h$ prediction has been
noted previously. Recent analyses include~\cite{Chanowitz:2008ix,Erler:2009jh,delAguila:2010mx}.} between
 the bound coming from direct searches for the Higgs, $M_h> 115\GeV$, and the electroweak fit~\cite{Chanowitz:2002cd}. 

It is beyond the scope of this analysis to perform a full fit to the electroweak data in this model. However, near the minimum of the fit the most substantial corrections are from the modification of the quark vertices in the $Z$ pole experiments due to $\Zp$ mixing\footnote{Since $M_Z$ is not significantly changed due to the compensation described above and since we are considering a leptophobic $Z'$, there is little
change in lepton vertices or the weak angle $\sin^2 \theta_W$.}.  We chose a subset of the EWPT observable most sensitive to this effect,
\be
\label{eqn:GammaZ}
\Gamma_Z &=& 2.4954 + \left( \frac{g_2}{g_1}\theta\right)f_{\Gamma_Z}\left(Q_Q,Q_{u^c},Q_{d^c}\right) \\
\label{eqn:Rl}
R_l &=& 20.735 + \left( \frac{g_2}{g_1}\theta\right) f_{R_l} \left(Q_Q,Q_{u^c},Q_{d^c}\right)  \\
\label{eqn:sigmaHad}
\sigma_{\rm Had} &=& 41.484 + \left( \frac{g_2}{g_1}\theta\right) f_{\sigma_{\rm had}} \left(Q_Q,Q_{u^c},Q_{d^c}\right)  \\
\label{eqn:Rb}
R_b &=& 0.21578 + \left( \frac{g_2}{g_1}\theta\right) f_{R_b}\left(Q_Q,Q_{u^c},Q_{d^c}\right)  \\
\label{eqn:AFBb}
A^b_{FB} &=& 0.1034 +  \left( \frac{g_2}{g_1}\theta\right) f_{A^b_{FB}}\left(Q_{Q_3},Q_{b^c}\right), 
\ee
where the functions $f\left(Q_Q,Q_{u^c},Q_{d^c}\right)$ are given by
\begin{eqnarray*}
f_{\Gamma_Z} &=& -0.31 \left( Q_{Q_1} +Q_{Q_2}\right) -1.7 Q_{Q_3}  \\ &-& 0.31 \left(Q_{d^c} + Q_{s^c}+ Q_{b^c}\right) + 0.65 \left( Q_{u^c} + Q_{c^c} \right) \\
f_{R_l}               &=& -3.7 \left( Q_{Q_1}+Q_{Q_2}\right) - 20.58 Q_{Q_3} \\ &-&3.73 \left(Q_{d^c} +Q_{s^c}\right) - 3.65 Q_{b^c}  + 7.69\left(Q_{u^c} +Q_{c^c}\right) \\
f_{\shad}     	&=& 2.90\left(Q_{Q_1}+ Q_{Q_2}\right) + 16.30 Q_{Q_3} \\&+& 2.96\left( Q_{d^c} +Q_{s^c}\right) + 2.89 Q_{b^c} - 6.09\left(Q_{u^c} +Q_{c^c}\right) \\
f_{R_b}              &=& 0.038\left(Q_{Q_1}+ Q_{Q_2}\right)-0.778 Q_{Q_3} \\&+&0.039\left(Q_{d^c}+Q_{s^c}\right) - 0.138 Q_{b^c}- 0.08\left(Q_{u^c} +Q_{c^c}\right) \\
f_{A^b_{FB}}		&=& -0.033 Q_{Q_3} + 0.18 Q_{b^c}.
\end{eqnarray*} 
We note that $A^b_{FB}$, where the largest discrepancy with the SM is present, is a function of only the third generation charges.

\subsection{Dijet Searches}
A $Z'$ which is produced via $q\bar q\rightarrow Z'$ can, of course, decay back into dijets.  Therefore, searches for dijet resonances 
set important limits on leptophobic models.  

\begin{figure}
\includegraphics[scale=0.7]{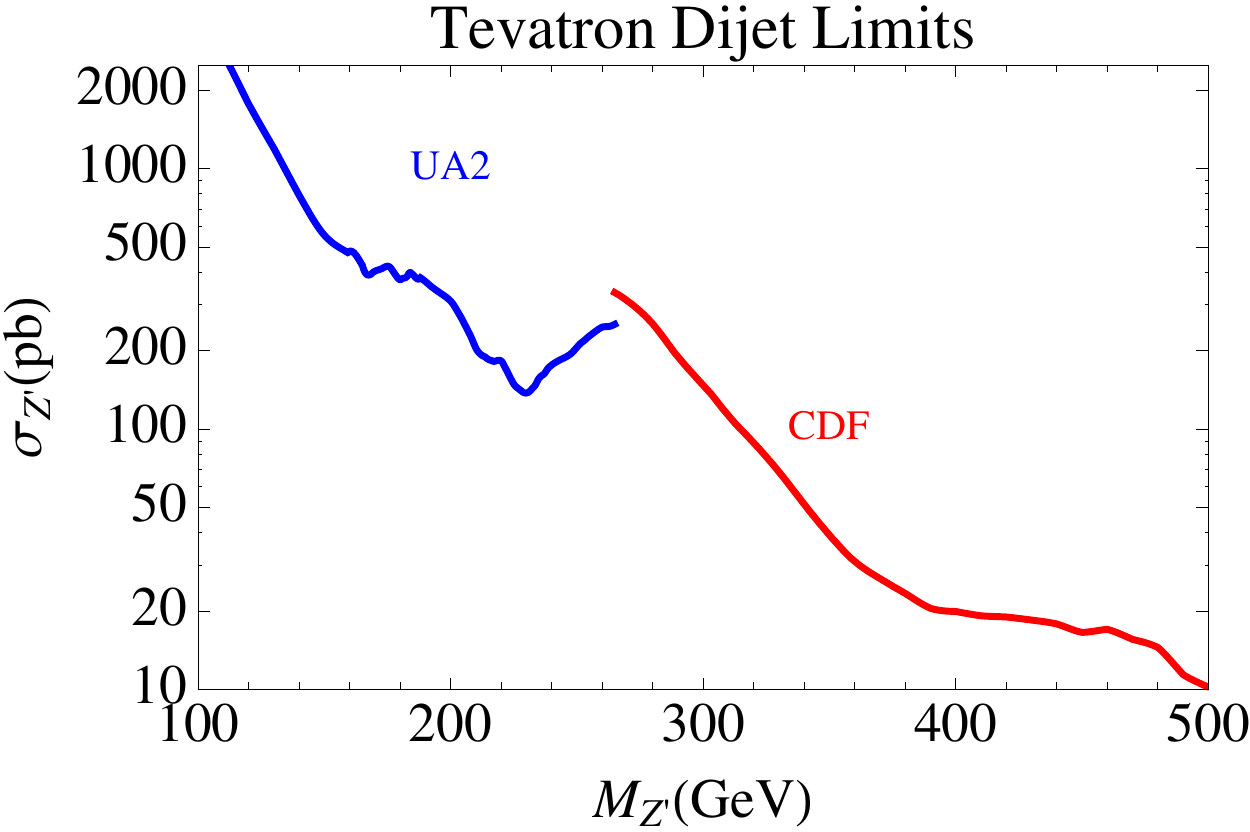}
\caption{Limits on the cross section times branching ratio into dijets for a $Z'$ produced at the Tevatron.  The 
low mass limits come from UA2, while those at high mass come from CDF.\label{fig:dijets}}
\end{figure}

The most stringent limits on a $Z'$ decaying into dijets near $M_{Z'}=270~{\rm GeV}$ actually come from a combination of UA2~\cite{Alitti:1993pn} and 
CDF results~\cite{Aaltonen:2008dn}, which we present in Fig.~\ref{fig:dijets}.  A $Z'$ produced at a rate of a few tens of picobarns (which is
typically
necessary for this sort of model to explain the CDF anomaly) 
is allowed across nearly the entire mass range, only running into serious constraints as $M_{Z'}\sim 500~{\rm GeV}$.  
It is amusing to note that the limits in the boundary region  between the two experiments are especially weak, allowing for a 
dijet production rate of over $250~{\rm pb}$ near $M_{Z'}=270~{\rm GeV}$.  While our models will not need 
nearly so large a coupling, it is comforting to know that this much freedom exists.


\section{$W+jj$ at CDF}
\label{sec:wjjcdf}


As discussed in the introduction, the CDF collaboration recently announced an intriguing excess in the dijet mass distribution between $120~{\rm GeV}<M_{jj}<160~{\rm GeV}$ taken from exclusive $W+2j$ samples where $W\rightarrow (e/\mu)+\nu$.  Their published data~\cite{Aaltonen:2011mk}, taken with $4.3~{\rm fb}^{-1}$, shows an excess of $3.2\sigma$ over the SM prediction and a recent~\cite{cdf73} updates using 
a larger $7.3~{\rm fb}^{-1}$ dataset increases this to $4.8\sigma$.  Here we will briefly discuss the properties of this excess, focusing primarily on those relevant to explanatory models with  
an $s$-channel topology and enhanced heavy flavor content.

While the CDF dijet excess could very well come from systematic errors in  background modeling, it is interesting to 
consider a new-physics explanation.  To this end, CDF estimates~\cite{Aaltonen:2011mk} that if the signal is coming from the 
associated production of a $150~{\rm GeV}$ particle along with a $W$ then the production rate times branching ratio into dijets must be of order 4pb.
While the efficiencies relevant for the $s$-channel production of $W+2j$ via a $Z'$ clearly differ from those of associated
production, in our simulations the difference did not amount to more than a ${\cal O}(10\%)$ effect.  Therefore, as the 4pb 
rate presented by CDF is only meant as a ballpark estimate, we will take it at face-value and apply it to our $s$-channel processes.

\begin{figure}
\begin{center}
\includegraphics[scale=0.65]{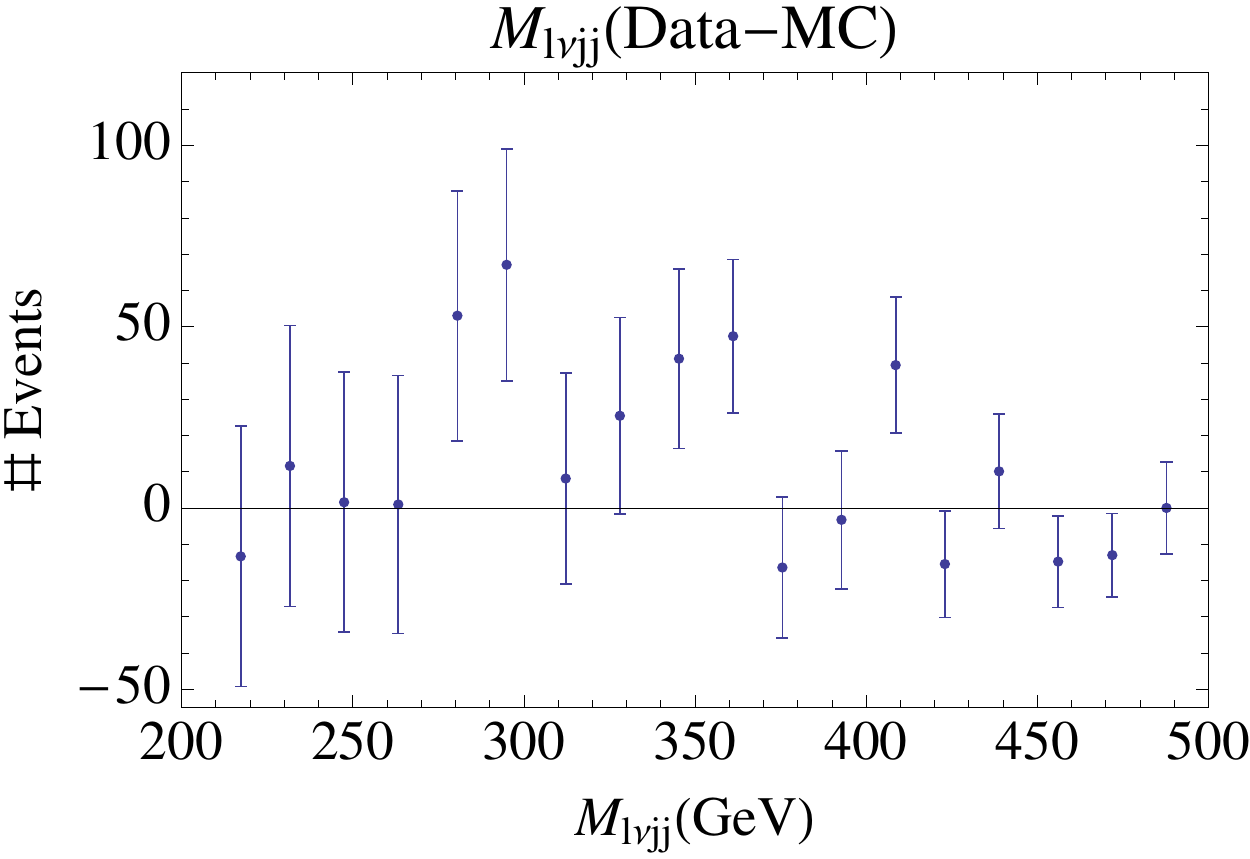}
\end{center}
\caption{The invariant mass of the $W+2j$ system for both $\mu$ and $e$ samples using the $4.3~{\rm fb}^{-1}$
CDF data.  Preliminary figures showing the distributions from the $7.3~{\rm fb}^{-1}$ dataset are available at~\cite{cdf73kin}.\label{fig:schannelmass}}
\end{figure}

\begin{figure}
\begin{center}
\includegraphics[scale=0.4]{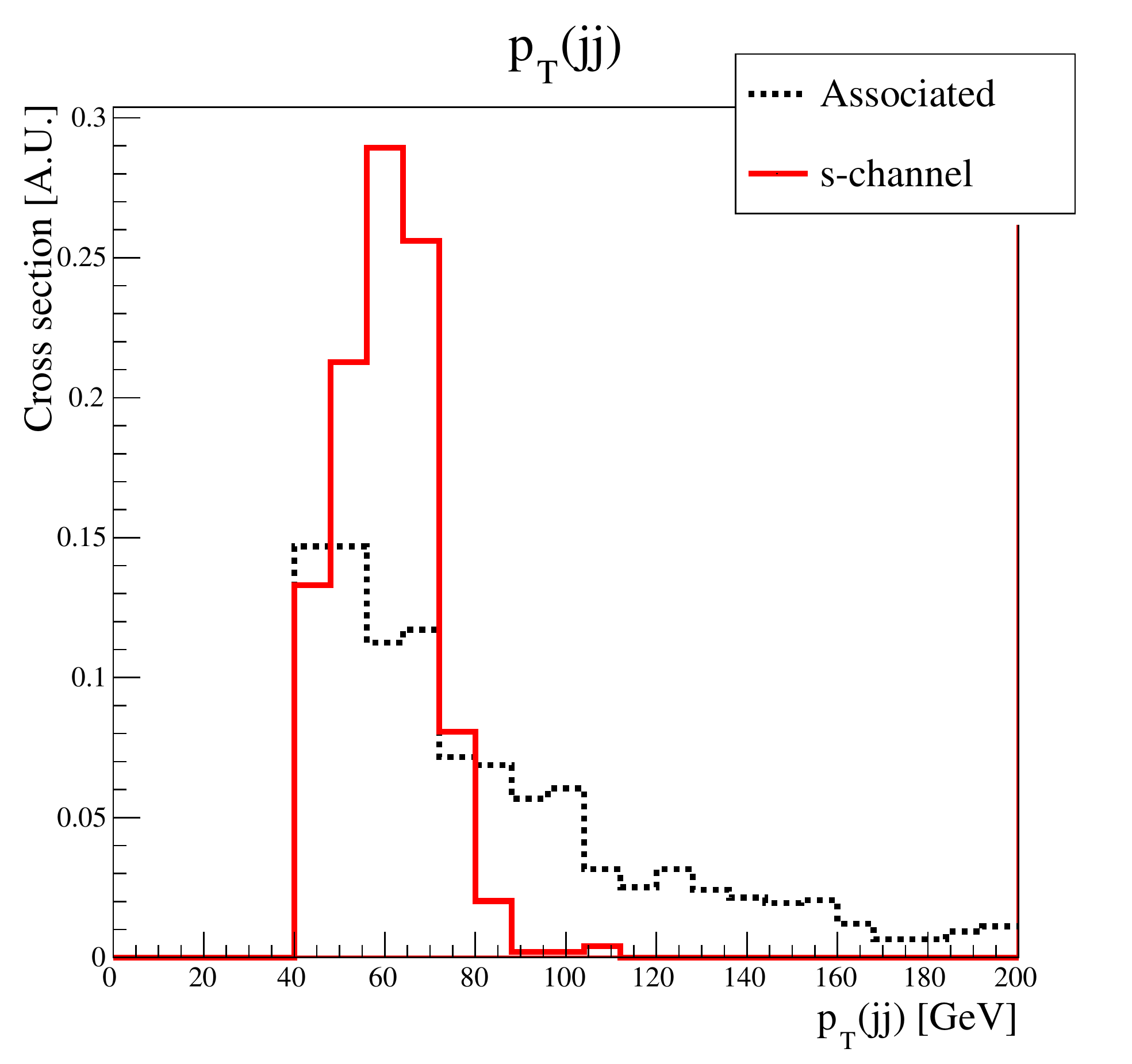}
\end{center}
\caption{The $p_T(jj)\sim p_T(W)$ distribution for an $s$-channel $Z'$ ($M_{Z'}=270~{\rm GeV}$, $M_{H^\pm}=150~{\rm GeV}$)
and a $Z'$ produced in association with a $W$ ($M_{Z'}=150~{\rm GeV}$).  These distributions were generated using the Fastjet~\cite{Cacciari:Fastjet,*Cacciari:2005hq}
implementation of JETCLU, operating on events generated by Madgraph 5~\cite{Alwall:2011uj} and showered in Pythia6~\cite{Sjostrand:2006za}.  The cuts applied were 
those used by CDF in their $W+jj$ analysis.\label{fig:wptzp}}
\end{figure}

Now, 
CDF has also considered the constraints placed on the excess from the distributions of various kinematic quantities.  The most
relevant for our purposes here is the invariant mass of the $M_{l\nu jj}$ system, which we present in Fig.~\ref{fig:schannelmass}.  
Unfortunately, it is difficult to derive any conclusions  from this plot - while there is a broad excess in the region 
surrounding $M_ {l\nu jj}\sim 270~{\rm GeV}$, the statistical error bars are far too large to make any concrete statements.  The
figure is included only to emphasize that $s$-channel physics is certainly still a viable explanation, despite lore to the contrary.  We 
note though that recent CDF preliminary results~\cite{cdf73kin} using $7.3~{\rm fb}^{-1}$, which employ more stringent cuts, are more suggestive 
of $s$-channel physics.  Furthermore, we note that the preliminary $p_T(W)$ distributions presented in~\cite{cdf73kin} seem 
to exhibit a sharp falloff near $p_T(W)\sim80~{\rm GeV}$ which may be more consistent with an $s$-channel process than with one in which a resonance is 
produced associatively (see Fig.~\ref{fig:wptzp}).  Thus, while CDF's current published results seem inconclusive at best, 
it might soon shed light on the production mechanism for the excess.

Finally, CDF provides some guidance on the flavor content of the two jets in the excess region.  
By performing an analysis looking at the $b$-tagging rate both in the signal 
region and in the immediate sidebands, the collaboration concludes
that there is no significant difference in  $b$-jet rates between the signal and the sidebands. However, 
while this is a strong result for the case of events with two $b$-tags, the statement becomes much weaker 
when only one is considered.  Indeed, the recent $7.3~{\rm fb}^{-1}$ results presented in \cite{cdf73kin}
show a small excess in the samples with a single $b$-tag.  Taken together, we estimate that they seem to fit a scenario where
the fraction of excess dijet events with a single $b$-quark is $\sim35\%\pm20\%$.  While it still seems
difficult for the dijet excess to be comprised entirely of single or double $b$-quarks, we note that a 
model yielding an increased concentration of $b$-quarks relative to the SM does seem compatible with the available data. 
We further note that the authors of Ref.~\cite{Hewett:2011nb}  came to similar conclusions in their analysis of the dijet heavy flavor content.


\section{High Energy Models}
\label{sec:hemodels}


We now move on to discuss three models with specific choices for the coupling of the $\Zp$ to the SM fields. The first two models we present suffer from serious tensions with at least one of the constraints mentioned above. Nevertheless, we believe they serve as instructive examples to manifest certain tensions that are present more generally in models of this type. They also make the charge choice of the third model evident. This last model serves as a viable explanation of the CDF anomaly and is in agreement with all other constraints.  

\subsection{Leptophobic}
\label{sec:leptoph}
The first model we consider is essentially the leptophobic $\Zp$ model of ref.~\cite{Agashe:1996dy, Georgi:1996ei}, with a slightly different choice for the charges. While it is  incapable of explaining the CDF excess without running into other constraints, it nicely illustrates some of the tensions we believe are present in many other possible leptophobic models. In this model both quarks and Higgs bosons are charged under the $\Uonep$. The Yukawa couplings are
\be
\lambda^{(u)}_{ij} \bar{Q}_i H_u u_j + \lambda^{(d)}_{ij}  \bar{Q}_i H_d d_j + \lambda^{(l)}_{ij} \bar{L}_i H_l e_j,
\ee
where $i$ and $j$ denote the three generations, and $H_l$ was introduced to avoid charging the leptons. If $\QHd=0$ then the two Higgses $H_d$ and $H_l$ can be identified. Since it is assumed that the Yukawa couplings are present above the scale of the $\Uonep$ breaking, they enforce three relations among the couplings $-\QQi + \QHu + \Qui=0$, $-\QQi + \QHd + \Qdi=0$, and $Q_{H_l} = 0$. 

Assuming family universal couplings, this model is anomalous. In ref.~\cite{Agashe:1996dy} the authors cancel the anomalies by adding 3 families of vector-like fermions under the SM gauge group\footnote{With 3 additional vector-like families, $SU(3)_c$ is no longer asymptotically free. It is fairly straightforward to avoid that by adding only a single family, however, then one must contend with non-integer charge assignments. Other choices are possible if one departs from the assumption of flavor universal couplings.}, dubbed $(Q'_L,Q'_R, u^{'c}_L,u^{'c}_R,d^{'c}_L,d^{'c}_R)$. This matter content then automatically leaves the SM gauge group anomaly free. The $\Uonep$ associated anomalies are then cancelled by choosing the right-labeled fields to have zero charge under $\Uonep$ and the left-labeled fields to have opposite charge compared with the SM fermions, $Q'_L = -Q_Q$, $u^{'c}_L = -u^c$, and $d^{'c}_L= -d^c$.

\begin{figure}[h]
\begin{center}
\includegraphics[width=0.4 \textwidth,height=0.2\textheight]{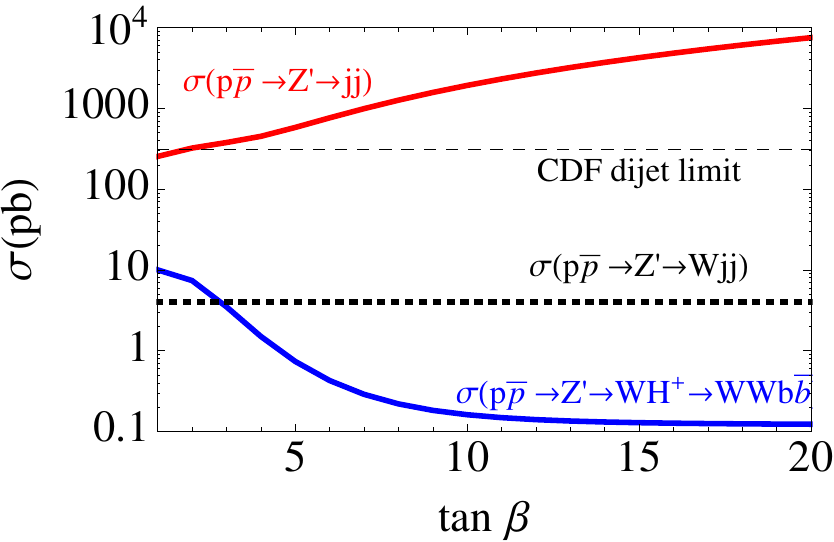}
\end{center}
\caption{Cross-section for the different decay modes of the $\Zp$ from Sec.~(\ref{sec:leptoph}) taken as a function of $\tan\beta$. The coupling, $g_2$, is normalized to yield a signal cross-section of 4 pb. The other parameters were chosen to be $\MZp = 270\GeV$, $\MHpm = 150\GeV$, $Q_d = -\QHd = 1$ with all other charges zero.}
\label{fig:ModelAConstraints}
\end{figure}

In Fig.~\ref{fig:ModelAConstraints} we show the cross-section for the different channels as a function of $\tan\beta$, where at every point we fixed $g_2$ to yield 4 pb for the $p\bar{p}\rightarrow Z' \rightarrow Wjj$ channel. The ratio of $\Zp\rightarrow H^\pm W^\mp$ to $\Zp\rightarrow jj$ in Eq.~\ref{eqn:WHjjRatio} makes it clear that it is difficult to have a sufficiently large signal without running afoul of the dijet constraints. It is possible to avoid these constraints for low $\tan\beta \lesssim 5$. However, one is then in conflict with EWPT. The combined pull on the SM observables is very large and the model results in a poor fit. The main difficulty is that a large $g_2$ is required to overcome the suppressed branching fraction $Z^\prime \to W^\mp H^\pm$ in the low $\tan\beta$ region. It is possible to relax this tension by requiring a lower cross-section of about 1 pb. However, it is not clear that one obtains enough events in the excess region with such a cross-section.   

\subsection{Higgsophilic I}

As we saw in the previous subsection, the main tension in generic leptophobic models is the large branching ratio of the $\Zp$ into dijets. To ameliorate this tension we now consider models where the quarks are uncharged under the $\Uonep$, with the possible exception of the right-handed top. Of course, some coupling to the first generation quarks must be introduced at some level or otherwise it is impossible to produce the $\Zp$ at the Tevatron at any appreciable rate. We achieve that by mixing the first generation quarks with a heavy vector-like set of quarks, as was recently discussed in ref.~\cite{Fox:2011qd}. So, henceforth, we assume that additional operators of the form
\be
M \bar{q}_L' q'_R + \lambda S \bar{q}_L' q_R 
\ee
are present, with $q_R = u_R$ or $d_R$, and where the $S$ field is the order parameter responsible for the breaking of $\Uonep$. The $q'$ fields form a vector representation under all gauge-groups and no anomalies are present. When the $S$ field obtains its VEV and the $\Uonep$ is broken, the $q_R'$ and $q_R$ fields mix. The lighter mass eigenstate is mostly a SM field $q_R^{(1)} =  q_R\cos\varphi  +q'_R \sin\varphi $ with $\sin\varphi = \lambda \langle S\rangle/\sqrt{M^2 +  \lambda^2 \langle S\rangle^2}$. Hence, this field inherits a small coupling to the $\Zp$, namely $Q_{q_R^{(1)}}= Q_{q'_R}~\sin^2\varphi $. This small charge naturally resolves the tension encountered with the previous model, where the partial width of $\Zp\rightarrow jj$ was overwhelmingly large. At the same time, a large enough production cross-section can be maintained by increasing the $\Uonep$ gauge coupling. Importantly, none of the conclusions below depends strongly on the details of the UV implementation, and in what follows we simply allow ourselves to choose the charge of the first generation quarks independently of the charge of the Higgses.

Since the top Yukawa is large we concentrate on scenarios where the operator responsible for the top mass is the usual SM one. One possibility is to charge the top quark under the $\Uonep$, 
\be
y_t \bar{Q}_3 H_u t_R + y_{ij}^{(u)} \frac{S}{\Lambda} \bar{Q}_i H_u u_{Rj} + y_{ij}^{(d)} \bar{Q}_i H_d d_{Rj},
\ee
with $\QHu = -Q_{t_R} = -Q_S$ and all other fields are uncharged. A small modification of this model would involve charging $H_d$ as well, but the resulting phenomenology is not markedly different. The main difficulty with this model is the contribution to $\delta\rho$,
\be
\delta \rho =  g_2^2 \QHu^2 \frac{v^2\sin^4\beta}{\MZp^2 - \MZZ^2}.
\ee
The only part of this expression that is potentially small is $g_2^2 Q_{H_u}^2$. This product can be expressed in terms of phenomenological quantities, such as the production cross-section and branching ratio, using Eqs.~(\ref{eqn:prodxsec}) and (\ref{eqn:WHjjRatio}),
\be
\delta\rho &\approx& \left(g_2^2 \sum_q Q_q^2\right) \frac{\QHu^2}{\sum_q Q_q^2}  \nonumber \\
&\approx& \frac{\sigma_{\Zp}}{\rm nb}\frac{12}{\sin^22\beta}\left(\frac{\Gamma_{W^\pm H^\mp}}{\Gamma_{jj}}\right).
\ee
Since we need a total cross-section of $\sigma_{\Zp} \gtrsim 10$ pb to explain the CDF anomaly, $\delta\rho \gg 10^{-3}$ even for moderate $\tan\beta$. This is a generic problem with Higgsophilic models where $\QHu\gtrsim \QHd$ since the contribution from the down-type Higgs is usually suppressed by $\tan^{-1}\beta$. This tension also points to the obvious resolution of this problem, namely models with $\QHu = 0$ where only the down-type Higgs is charged under the $\Zp$. 

\subsection{Higgsophilic II}

\begin{figure}[h]
\begin{center}
\includegraphics[width=0.4 \textwidth,height=0.2\textheight]{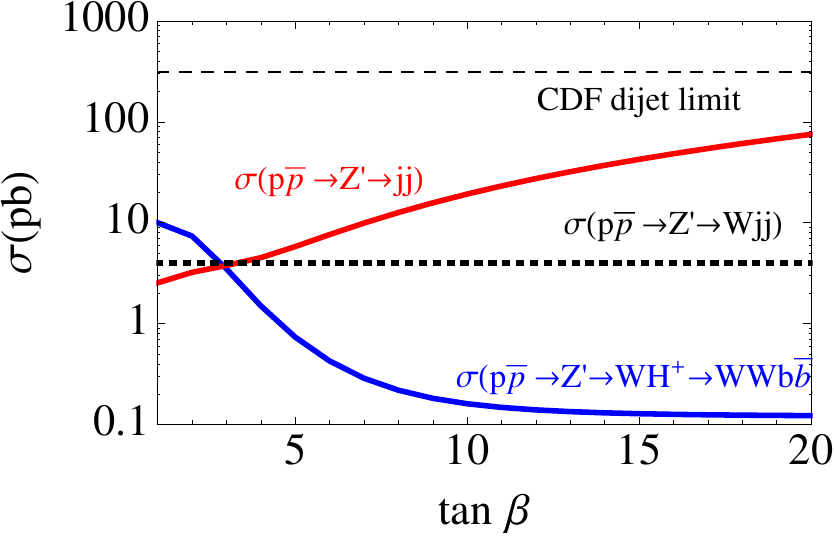}
\end{center}
\caption{Cross-section for the different decay modes of the $\Zp$ as a function of $\tan\beta$. The coupling, $g_2$, is normalized to yield a signal cross-section of 4 pb. The other parameters were chosen to be $\MZp = 270\GeV$, $\MHpm = 150\GeV$, $\QHd = 1$, $Q_u = 0.1$ with all other charges zero.}
\label{fig:ModelCConstraints}
\end{figure}

The tensions discussed above point to a very simple model where the only field charged under the $\Uonep$ is the down-type Higgs, $H_d$. In this case, the Yukawa couplings to matter must follow 
\be
 y_{ij}^{(u)}  \bar{Q}_i H_u u_{Rj} + y_{ij}^{(d)} \frac{S}{\Lambda} \bar{Q}_i H_d d_{Rj},
\ee
with $Q_S = -\QHd$. A similar operator can be written for the leptons. In Fig.~\ref{fig:ModelCConstraints} we plot the cross-section for each of the channels as a function of $\tan\beta$, again normalizing the signal cross-section to 4 pb, and the up-type charge $Q_{u^c} = 0.1$. This plot suggests that the model is allowed for all values of $\tan\beta$. With this normalization the $\Uonep$ coupling is sizable, $g_2\sim 1$ for $\tan\beta = 5$ and increases further for larger $\tan\beta$ as the partial width of $\Zp\rightarrow W^\pm H^\mp$ decreases. While this value does depend on the choice of normalization for the signal cross-section and effective quark charge, we note that the $\mathcal{O}(1)$ coupling encountered here is a fairly generic feature of this model~\footnote{This point is also independently emphasized in ref.~\cite{nima}.}.  

\begin{figure}[h]
\begin{center}
\includegraphics[width=0.4 \textwidth,height=0.2\textheight]{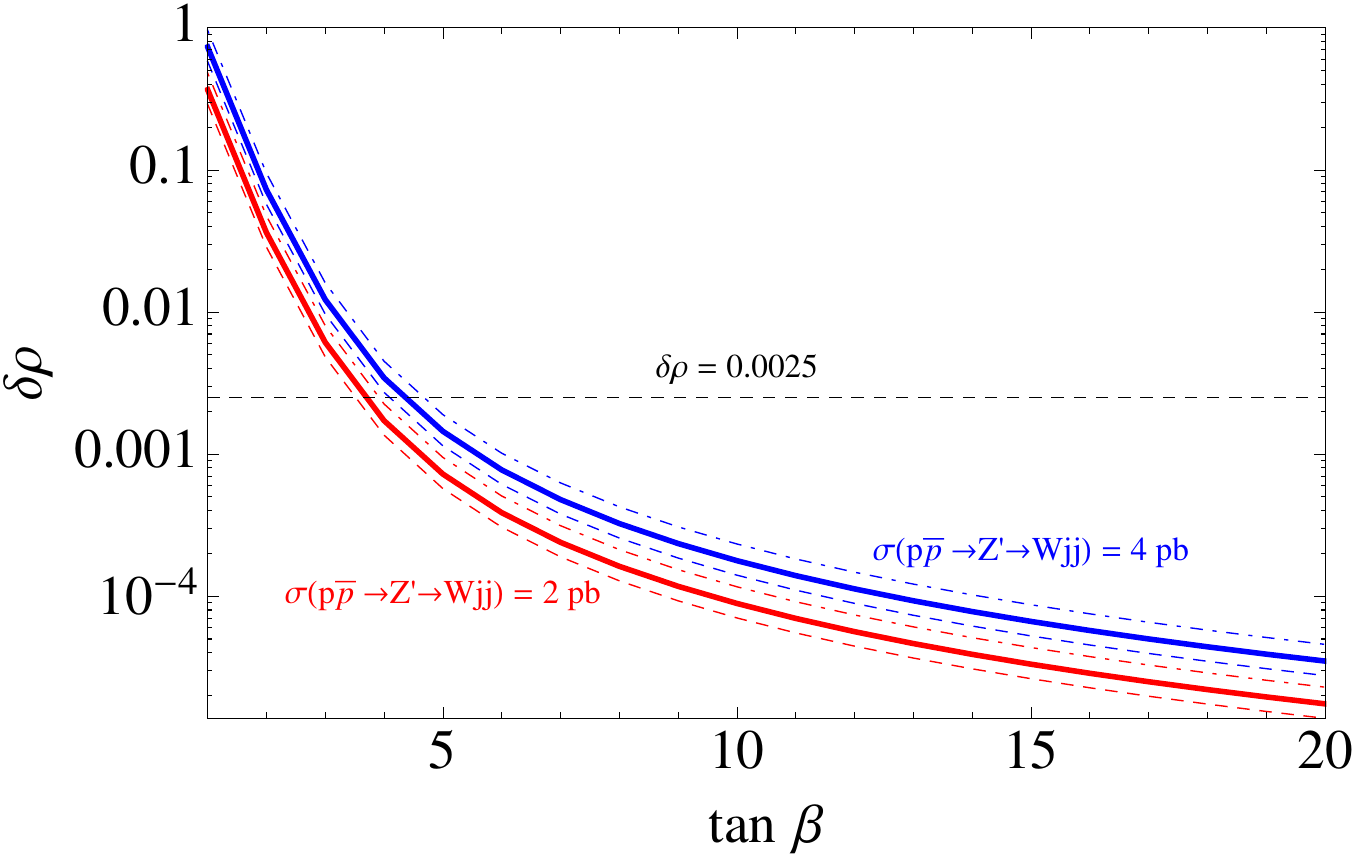}
\end{center}
\caption{The upper curves in blue show $\delta\rho$ vs. $\tan\beta$ for the Higgsophilic model discussed in the text, where the coupling, $g_2$, is normalized to yield a signal cross-section of 4 pb for $\MZp = 240\GeV$ (dot-dashed), $\MZp = 270\GeV$ (thick), and $\MZp=300\GeV$ (dashed). The other parameters were $\MHpm = 150\GeV$, $\QHd = 1$, and an effective up-type quark coupling of $Q_u = 0.1$ with all other charges zero. The lower red curves depict the same thing with $g_2$ now normalized for a 2 pb signal cross-section.}
\label{fig:deltarho_vs_tanbeta}
\end{figure}

\begin{figure}[h]
\begin{center}
\includegraphics[width=0.4 \textwidth,height=0.2\textheight]{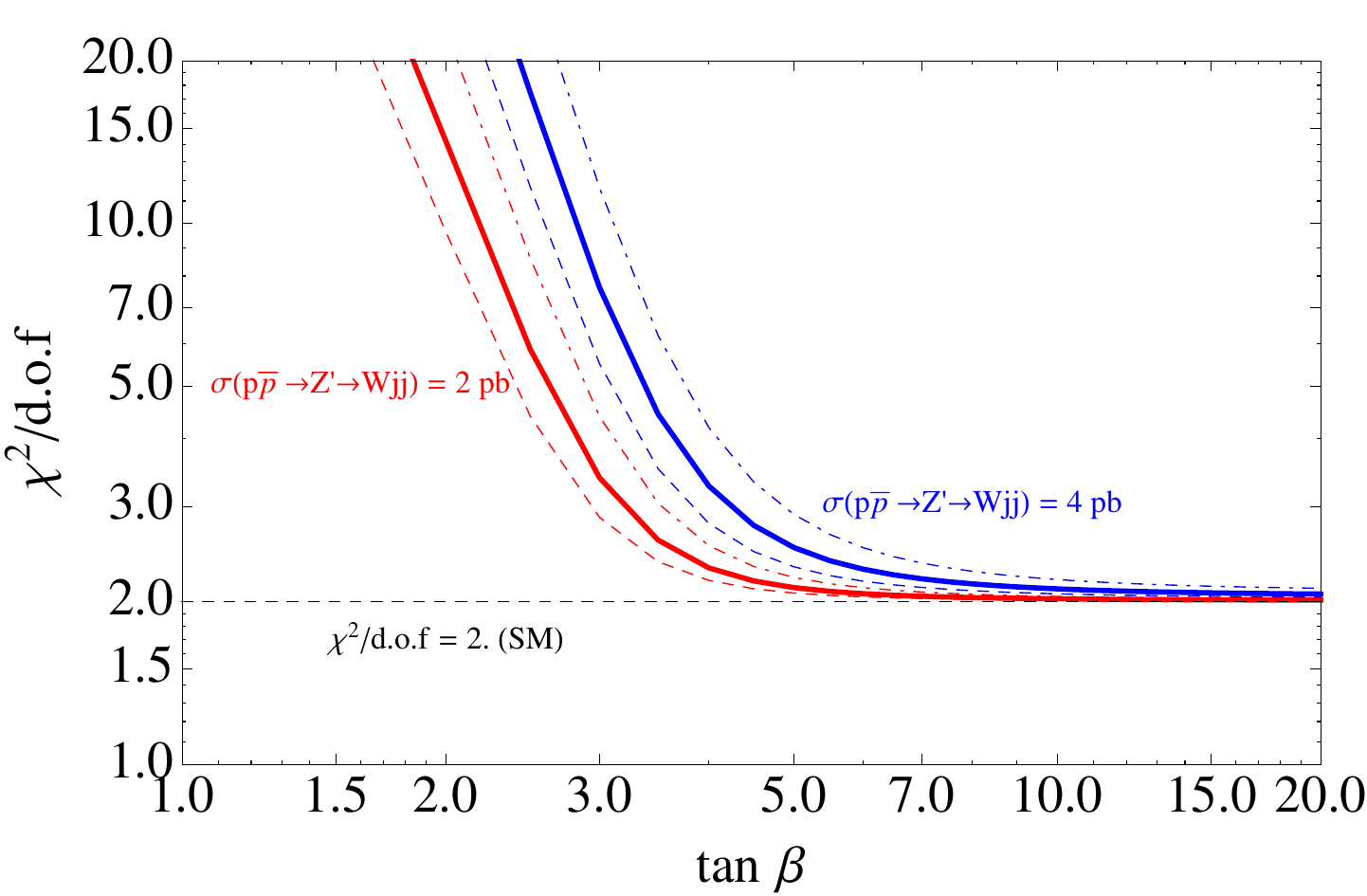}
\end{center}
\caption{The upper curves in blue show $\chi^2$ vs. $\tan\beta$ for the Higgsophilic model discussed in the text, where the coupling, $g_2$, is normalized to yield a signal cross-section of 4 pb for $\MZp = 240\GeV$ (dot-dashed), $\MZp = 270\GeV$ (thick), and $\MZp=300\GeV$ (dashed). The other parameters were $\MHpm = 150\GeV$, $\QHd = 1$, and an effective up-type quark coupling of $Q_u = 0.1$ with all other charges zero. The lower red curves depict the same thing with $g_2$ now normalized for a 2 pb signal cross-section. }
\label{fig:Chi2_vs_tanbeta}
\end{figure}

The contribution to the $\delta\rho$ parameter is suppressed by $\cos^4\beta$ and is easily within present limits for moderate $\tan\beta$, as we show in Fig.~\ref{fig:deltarho_vs_tanbeta}. In fact,  as discussed above, this contribution can help resolve the tension between the direct and indirect searches for the Higgs boson and allow for a higher Higgs mass in the EW fits. In Fig.~\ref{fig:Chi2_vs_tanbeta} we show the resulting $\chi^2$ for the observables in Eqs.~(\ref{eqn:GammaZ}-\ref{eqn:AFBb}) as compared with the SM. Again, for moderate values of $\tan\beta$ there is excellent agreement with EW precision data. 

At moderate $\tan\beta$ the charged Higgs decay is dominated by decay into bottom-charm or strange-charm pairs. As shown in Fig.~\ref{fig:HpmPW}, the partial width quickly asymptotes to approximately $BR(H^+\rightarrow c\bar{b}) = 70\%$ and  $BR(H^+\rightarrow c\bar{s}) = 30\%$. This means that we expect a sizable concentration of single bottom content in the dijets associated with the $\approx 150\GeV$ resonance seen by CDF. Whether this prediction is born in data remains to be seen. 

\begin{figure}[h]
\begin{center}
\includegraphics[width=0.4 \textwidth,height=0.2\textheight]{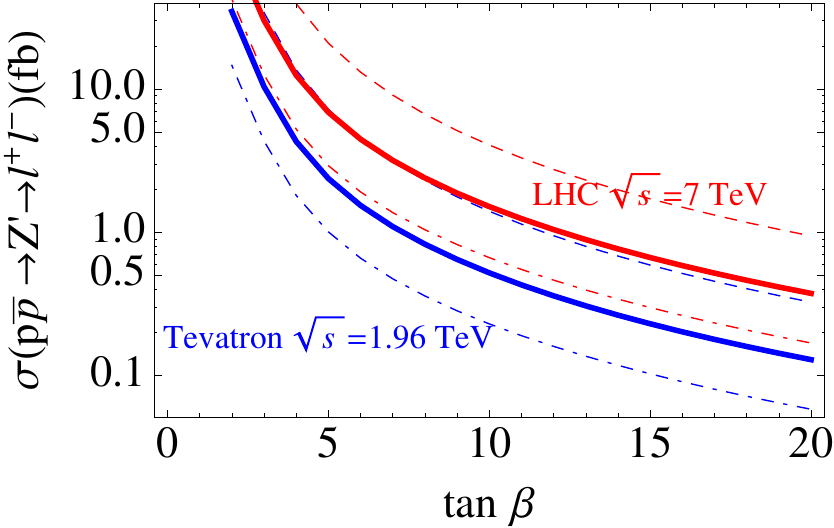}
\end{center}
\caption{The lower curves in blue show the predicted $\sigma(p\bar{p}\rightarrow \Zp) \times {\rm BR}(\Zp\rightarrow l^+l^-)$ at the Tevatron, for $\MZp = 240, 270, 300\GeV$ (dot-dashed, solid, dashed). The upper curves show the corresponding rates for $pp$ at the LHC with $\sqrt{s} = 7\TeV$.}
\label{fig:ZpLeptonXS}
\end{figure}

Finally, the mixing with the SM $\ZZ$ inevitably results in a decay into a di-lepton pair, Eq.~(\ref{eqn:PWlep}). In Fig.~\ref{fig:ZpLeptonXS} we plot the rate for production of $\Zp\rightarrow l^+l^-$ at the Tevatron as well as the LHC for each generation, i.e., $l^\pm = e^\pm$, $\mu^\pm$, $\tau^\pm$. For $\tan\beta \lesssim 10$ this channel should become visible at the Tevatron right about now. This is a fairly robust prediction of the model and calls for a dedicated search for a resonance at $\MZp\sim 270\GeV$ in the di-lepton channel.

\section{Discussion}
\label{sec:discussion}


The CDF collaboration recently presented an excess in the dijet invariant mass distribution 
taken from events in the $W+jj$ exclusive sample.  While the excess may well be a systematic error in background modeling, we believe it 
is still useful to consider new-physics explanations of this anomaly as they motivate a careful study of other kinematical distributions involving different systamatics.  

Here we have presented a model which attributes the dijet invariant mass excess to the decay of a charged Higgs 
produced via an $s$-channel $Z'$: $pp\rightarrow \Zp\rightarrow W^\pm H^\mp$ where $\MZp \sim 270\GeV$ and $M_{H^\pm}\sim 150\GeV$.  In our model the leptons and up-type Higgs are uncharged 
under the new $U(1)'$ group, while the quarks only acquire a small charge via mixing with other heavy states.  As the down-type Higgs doublet is the only field at the 
electroweak scale charged under the $U(1)'$, the model is both Higgsophilic and leptophobic.  Thus, the $\Zp$ model we present is able to comfortably explain the excess in $W+jj$ events without running into tension with either precision electroweak data or collider bounds. In fact, through its contribution to the $\delta\rho$ parameter, it may even resolve the considerable tension between direct and indirect searches for the Higgs boson.

Of course, the main motivation behind the construction of new-physics models to explain anomalies is that they give 
predictions visible in distributions subject to different systematics.  We therefore emphasize that the $\Zp$ model we present specifically predicts 
that the 
events in the $M_{jj}$ excess region will show (1) a bump near 270 GeV in the
distribution of $M_{l \nu j j }$, (2) an edge in the $W$ $p_T$ distribution  near $\sim 80~{\rm GeV}$, (3) an increased heavy-flavor concentration coming from $H^\pm\rightarrow bc$, and (4) a small, but possibly observable resonance in the di-lepton distribution near $M_{l^+l^-}\sim 270\GeV$ as a result of the $\ZZ/\Zp$ mixing.  

Finally, we note that we have not considered in detail the neutral Higgs sector of this model, although we do not believe that these considerations will change any of our conclusions thus far.
In fact, it would be interesting to pursue this sort of model building in future work as the neutral Higgs sector can yield additional interesting signatures of the $\Zp$. One might hope to see the decay of the $\Zp$ into a $\ZZ$ and a neutral Higgs, where $h\rightarrow WW^\ast$ or $H\rightarrow b \bar b$. It may also be worthwhile to consider a supersymmetric version of this setup as it may offer further constraints on the theoretical framework.



\begin{acknowledgments}

We would like to thank Nima Arkani-Hamed, Pierluigi Catastini, Csaba Csaki, Adam Martin, Matt Reece, and Matt Schwartz for useful discussions.
J.F. is supported by the NSF under grant PHY-0756966. D.K. is supported by a Simons postdoctoral fellowship and by an LHC-TI travel grant. The work of P.L. is supported by an IBM Einstein Fellowship and by NSF grant PHYÐ0969448. I.Y. is supported by the James Arthur fellowship.

\end{acknowledgments}


\appendix
\renewcommand{\theequation}{A-\arabic{equation}}
\setcounter{equation}{0}

\section{3-body decays of the charged Higgs}


In this section, we present the formula of the 3-body partial width of the charged Higgs in Eq.(\ref{eqn:HpmTObbW}).
\be
&&\frac{\Gamma\left(H^{\pm} \rightarrow W^\pm b\bar{b} \right)}{\MHpm} = \frac{3 }{128\pi^3 v^4 \tan^2\beta} \nonumber \\
&& \left(m_t^4 f_L\left(\kappa_t,\kappa_W\right)+\MHpm^2m_b^2\tan^4\beta f_R\left(\kappa_t,\kappa_W\right) \right), \nonumber 
\ee
with
\be
&&f_L\left(\kappa_t,\kappa_W\right)=\frac{\kappa_W^2}{\kappa_t^3}(4\kappa_W\kappa_t+3\kappa_t-4\kappa_W)\log\frac{\kappa_W \left(\kappa _t-1\right)}{\left(\kappa _t-\kappa _W\right) } \nonumber \\
 &&+(3\kappa_t^2-4\kappa_t-3\kappa_W^2+1)\log\frac{ \kappa _t-1}{\kappa _t-\kappa _W} -\frac{5}{2}\nonumber \\
 &&+\frac{\kappa_W-1}{\kappa_t^2}\left(-3\kappa_t^3+\kappa_W\kappa_t+2\kappa_W^2\kappa_t-4\kappa_W^2\right)\nonumber \\
 &&+\kappa_W\left(4-\frac{3}{2}\kappa_W\right) \\
   &&f_R\left(\kappa_t,\kappa_W\right)=  -2\frac{\kappa _W^3}{\kappa_t^2}\log\frac{\kappa_W \left(\kappa _t-1\right)} {\left(\kappa _t-\kappa_W\right)} \nonumber \\
 &&  +2\left(2 \kappa _t^3-3 \kappa_t^2-\kappa _t \left(3 \kappa _W^2-1\right)+\kappa _W^2 \left(\kappa _W+3\right)\right) \log\frac{\kappa _t-1} {\kappa _t-\kappa_W} \nonumber \\
&&+2\left(1-\kappa _W\right)\left(\frac{\kappa _W^2}{\kappa _t}+2\kappa_t^2+\kappa_t(\kappa_W-2)\right) \nonumber \\
 &&  +\frac{1}{3} \left(14 \kappa _W^3-9 \kappa _W^2-6 \kappa _W+1\right)
\ee
where $\kappa_t = m_t^2/\MHpm^2$ and $\kappa_W=M_W^2/\MHpm^2$. 


\bibliography{leptophobicZ'-Wjj}
\end{document}